\documentclass[pra,twocolumn,letterpaper,amsfonts,floatfix]{revtex4}%

\usepackage[dvips]{graphicx}
\usepackage{amsmath}

\usepackage{amssymb}

\newcommand{\Z}{\mathbb{Z}}

\newcommand{\drawbox}[3][0pt]{\raisebox{#1}%
{\setlength{\fboxsep}{0pt}\framebox{\rule{#2}{0pt}\rule{0pt}{#3}}}}

\begin{document}

\title{Refinement trajectory and determination of eigenstates by a wavelet
based adaptive method}

\author{J\'anos Pipek}
\affiliation{Department of Theoretical Physics, Institute of Physics,\\
             Budapest University of Technology and Economics,\\
             H--1521 Budapest, Hungary\\}
\author{Szilvia Nagy}
\affiliation{Department of Telecommunication,\\
             Jedlik \'Anyos Institute of Informatics, Electrical and Mechanical Engineering,\\
             Sz\'echenyi Istv\'an University, H--9026 Gy\H{o}r, Egyetem t\'er 1, Hungary\\}

%\date{}

\begin{abstract}
The detail structure of the wave function is analyzed at various refinement levels using
the methods of wavelet analysis. The eigenvalue problem of a model system is solved in
granular Hilbert spaces, and the trajectory of the eigenstates is traced in terms of the
resolution. An adaptive method is developed for identifying the fine structure localization
regions, where further refinement of the wave function is necessary.
\end{abstract}

\maketitle

\section{Introduction}

Multiresolution or wavelet analysis (MRA) \cite{Daub,Chui} is
increasingly used in advanced data compression and image coding
algorithms \cite{JPEG}, and several promising approaches applied
this technique in describing electronic structures. Arias \emph{et
al.} \cite{AriasRMP} elaborated a Kohn-Sham equation based
\emph{density functional} (DFT) method, Goedecker and Ivanov
\cite{goed} have also successfully used wavelets for the solution
of Poisson's equation. There are also attempts to extend wavelet
calculations to three dimensional structures in a computationally
manageable manner \cite{Beylkin1}.

The basic principle of the application of MRA is the recognition
that the details of the electronic structure are not distributed
equally over different parts of the system. In data compression
the details of the picture are systematically included by
consecutive refinements of the image only in those spatial regions
where it is necessary. A similar approach to the electronic wave
function suggests that the fine details of the distribution are
concentrated around the nuclear cusps and singularities of the
two-electron density matrix. It has been shown, that the
surroundings of a molecule can be described at a rather rough
resolution level \cite{negy}. We have also demonstrated
\cite{ket}, that the extremely hardly representable
electron-electron cusp singularity of the two-electron density
operator can be easily reproduced by the method of multiresolution
analysis.

The applicability of these approaches strongly depends on the
necessary depth of refinement. Clearly, a uniform refinement of
the distributions leads to an exponential increase of the
computational efforts with growing resolution level. The region
where the refinements are applied has to be confined, in order to
achieve computationally manageable algorithms. In this
contribution we study some questions concerning several aspects
raised by the above considerations. We analyze the detail
structure of the wave function, we explore how the spatial regions
where further refinement is necessary are to be identified, and
how can the extension of the eigenvalue problem to new subspaces
be restricted. We trace how the projection of the wave function to
different resolution level subspaces is changing in the course of
the consecutive refinement steps. The performance of the adaptive
refinement algorithm is tested by numerical comparisons to the
exact analytical solution of the harmonic oscillator.

\section{Resolution structure of the Hilbert space}\label{sec:MRA}

Multiresolution analysis theory ensures that an arbitrary element
of the Hilbert space $\Psi\in\mathcal{H}$ can be exactly
decomposed into orthogonal components as
\begin{equation}\label{psidecomp}
  \Psi(x)=\sum_{\ell\in\mathbb{Z}} c_\ell\; s_{0 \ell}(x)+\sum_{\rule{0pt}{1.11ex}m=0}^\infty\sum_{\ell\in\Z}d_{m \ell}
  \;w_{m\ell}(x).
\end{equation}
Here, the basis functions $s_{0\ell}(x)=s_0(x-\ell)$ are
equidistant shifts of the ``mother scaling function'' $s_0$ over a
coarse grid of spacing 1. Refinements to the first approximation
described by the first summation in (\ref{psidecomp}) are
introduced by the $m=0$ level wavelets $w_{0\ell}(x)=w_0(x-\ell)$.
The ``mother wavelet'' $w_0$ is orthogonal to $s_0$, and
generally, $\langle s_{0\ell}|s_{0k}\rangle=\delta_{\ell k}$,
$\langle s_{0\ell}|w_{0k}\rangle=0$ and $\langle
w_{0\ell}|w_{0k}\rangle=\delta_{\ell k}$. Further refinements are
expanded by the basis vectors $w_{m\ell}(x)=2^{m/2}w_0(2^m
x-\ell)$ for resolution levels $m=1,2,\ldots$, with the
orthogonality relations $\langle s_{0\ell}|w_{mk}\rangle=0$ and
$\langle w_{m\ell}|w_{j\,k}\rangle=\delta_{m j}\,\delta_{\ell\,
k}$. The subspaces $V_0=\mbox{span} \{s_{0 \ell}|\,\ell\in\Z\}$,
$W_m=\mbox{span} \{w_{m\ell}|\,\ell\in\Z\}$ constitute a complete
decomposition of the Hilbert space
\begin{equation}\label{VWdecomp}
  \mathcal{H}=V_0\oplus W_0\oplus W_1\oplus\cdots,
\end{equation}
i.e., expansion (\ref{psidecomp}) is exact.

We define the resolution structure of the wave function by the
series of components
\begin{eqnarray}\label{ResStruct}
    P_0\Psi &=& \sum_{\ell\in\mathbb{Z}} c_\ell\; s_{0 \ell} \nonumber\\
    Q_0\Psi &=& \sum_{\ell\in\mathbb{Z}} d_{0\ell}\; w_{0 \ell} \nonumber\\
    Q_1\Psi &=& \sum_{\ell\in\mathbb{Z}} d_{1\ell}\; w_{1 \ell}, \\
    &\vdots& \nonumber
\end{eqnarray}
where $P_0$ and $Q_m$ are orthogonal projection operators to
subspaces $V_0$ and $W_m$, respectively. As according to
Parseval's equality
\begin{equation}\label{Parseval}
  \|P_0\Psi\|^2+\sum_{m=0}^\infty \|Q_m\Psi\|^2=1,
\end{equation}
we can measure the $m$th level complexity of the wave function by
the number $\|Q_m\Psi\|^2$, which characterizes how important is
the detail space $W_m$ in expanding $\Psi$.

The decision, to include or omit $W_m$ is crucial in developing
useful algorithms for the following reasons. The basis functions
of $W_m$ are ``sitting'' on an equidistant grid with a grid length
of $\sim 2^{-m}$. If a function constrained to a finite domain of
the space is expanded, the number of basis functions in $W_m$ is
increasing like $\sim 2^{mD}$, where $D$ is the dimension of the
system. This exponential ``explosion'' makes the direct
application of (\ref{psidecomp}) unacceptable. Detail spaces with
negligible $\|Q_m\Psi\|^2$ (lower than a predefined threshold) can
be completely ignored. The experience \cite{negy,Dahmen} of
finding details in a constrained region of the space leads,
however, to the conclusion, that even if
\begin{equation}
\label{significant}
 \|Q_m\Psi\|^2=\sum_{\ell\in\Z}d_{m\ell}^2
\end{equation}
is significant, only very few terms in the summation contribute
essentially to its value. This recognition would help to avoid the
exponential explosion mentioned above, by using the restricted
detail space $\widetilde{W}_m=\mbox{span}\{w_{m\ell}|\mbox{ where
}d_{m\ell}\mbox{ is significant}\}$.

The outlined strategy is, however, useless, if the choice of the
significant $d_{m\ell}$ would be based on a prior calculation of
\emph{all} coefficients, and testing how many $d_{m\ell}$ are
necessary to fulfill (\ref{significant}) with a good
approximation. Clearly, a predictive method is needed.

\subsection{Decomposition of exact wave functions}

\begin{figure}[h!]
  \setlength{\unitlength}{\textwidth}
  \begin{picture}(0.47,0.39)
    \put(0.009,0.02){\includegraphics[width=0.45\textwidth,height=0.35\textwidth]{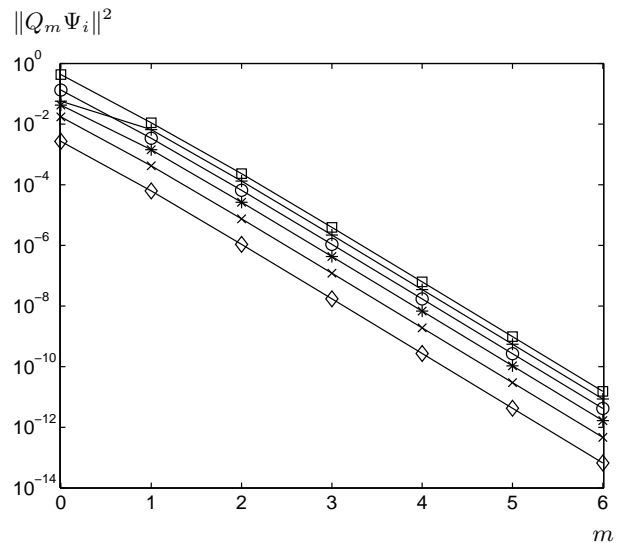}}%
    \put(0.009,0.38){\small $\|Q_m\Psi_i\|^2$}
    \put(0.439,0.00){\small $m$}
  \end{picture}
\caption{\label{fig:1} The amplitudes of the projections of the exact eigenfunctions of the
Hamiltonian (\ref{HO_Hamiltonian}) with $\omega=1$ to the detail spaces $W_m$ for
$m=0,\ldots,6$. The sign $\lozenge$ stands for the ground state $i=0$, whereas the signs
$\times$, $\ast$, $\circ$, $+$ and \drawbox{1ex}{1ex} denote the excitations $i=1,2,3,4,5$,
respectively. Atomic units are used.}
\end{figure}

We will illustrate the above consideration with a simple exactly
solvable example in $D=1$ dimension. The case of $D=3$ is expected
to have similar behavior with correspondingly greater numbers of
basis functions. The ground state and some excited states of the
standard example of the 1D harmonic oscillator with the
Hamiltonian
\begin{equation}
\label{HO_Hamiltonian}
 H=-\frac{1}{2}\nabla^2+\frac{\,\omega^2\!}{2}\, x^2
\end{equation}
are analyzed. One reason for choosing this system is that the
matrix elements $\langle s_{0\ell}|H|s_{0k}\rangle$, $\langle
s_{0\ell}|H|w_{jk}\rangle$ and $\langle w_{m\ell}|H|w_{jk}\rangle$
can be calculated exactly for (\ref{HO_Hamiltonian}), making
possible to avoid inaccuracies not connected to Hilbert space
constraints. The grid length of the scaling function subspace
$V_0$ was set to 1~a.u. The expansion coefficients were calculated
as
\begin{equation}\label{expcoeff}
c_\ell=\langle s_{0\ell}|\Psi\rangle\qquad\mbox{and}\qquad
 d_{m\ell}=\langle w_{m \ell}|\Psi\rangle,
\end{equation}
where $\Psi$ stands for the exact ground or excited state wave function. The Daubechies-6
\cite{Daub} set was chosen for the multiresolution basis set $\{s_{0\ell},w_{m\ell}\}$. The
elements of this set have finite support (the mother scaling function is zero outside of
the interval $[0,5)$), and both the scaling functions and the wavelets are differentiable.
The scalar products in (\ref{expcoeff}) were numerically calculated on a grid of
$2^{-15}$~a.u.\ density. Fig.~\ref{fig:1} shows the values $\|Q_m\Psi_i\|^2$ for various
excitations $i=0,\ldots,5$ with $\omega=1$, whereas the projections to the scaling function
subspace are given as follows: $\|P_0\Psi_0\|^2=0.9972$, $\|P_0\Psi_1\|^2=0.9822$,
$\|P_0\Psi_2\|^2=0.9561$, $\|P_0\Psi_3\|^2=0.8637$, $\|P_0\Psi_4\|^2=0.9362$ and
$\|P_0\Psi_5\|^2=0.5582$.
It is clear, that the bulk of the states is described at level $m=0$, and the contributions
of the detail spaces disappear with increasing $m$. It is important to realize, that even
if the number of basis functions in subspace $W_m$ increases as $\sim 2^m$, the \emph{total
contribution} of the subspace to the wave function exponentially decreases. This results
coincides with our previous statement \cite{egy} that many-electron density operators can
not contain arbitrary fine (nor rough) details.

\subsection{Fine structure localization}

Besides a ``vertical'' truncation of the Hilbert space over a
sufficiently large resolution level $M$, there is a possibility of
reducing the size of the subspaces $W_m$ in a ``horizontal''
truncation process, by decimating the basis functions $w_{m\ell}$
which belong to those spatial regions where the wave function does
not contain fine details.

In order to study the extent of this fine structure localization, we have also examined,
how many coefficients $c_\ell$ and $d_{m\ell}$ are essential in the norms $\|P_0\Psi_i\|^2$
and $\|Q_m\Psi_i\|^2$, respectively. The projections of the wave function to the restricted
subspaces $\widetilde{V}_0$ and $\widetilde{W}_m$ are defined by the projectors
$\widetilde{P}_0$ and $\widetilde{Q}_m$.

\begin{figure*}[h!]
  \setlength{\unitlength}{\textwidth}
  \begin{picture}(1,0.8)
    \put(0.039,0.02){\includegraphics[width=0.45\textwidth,height=0.35\textwidth]{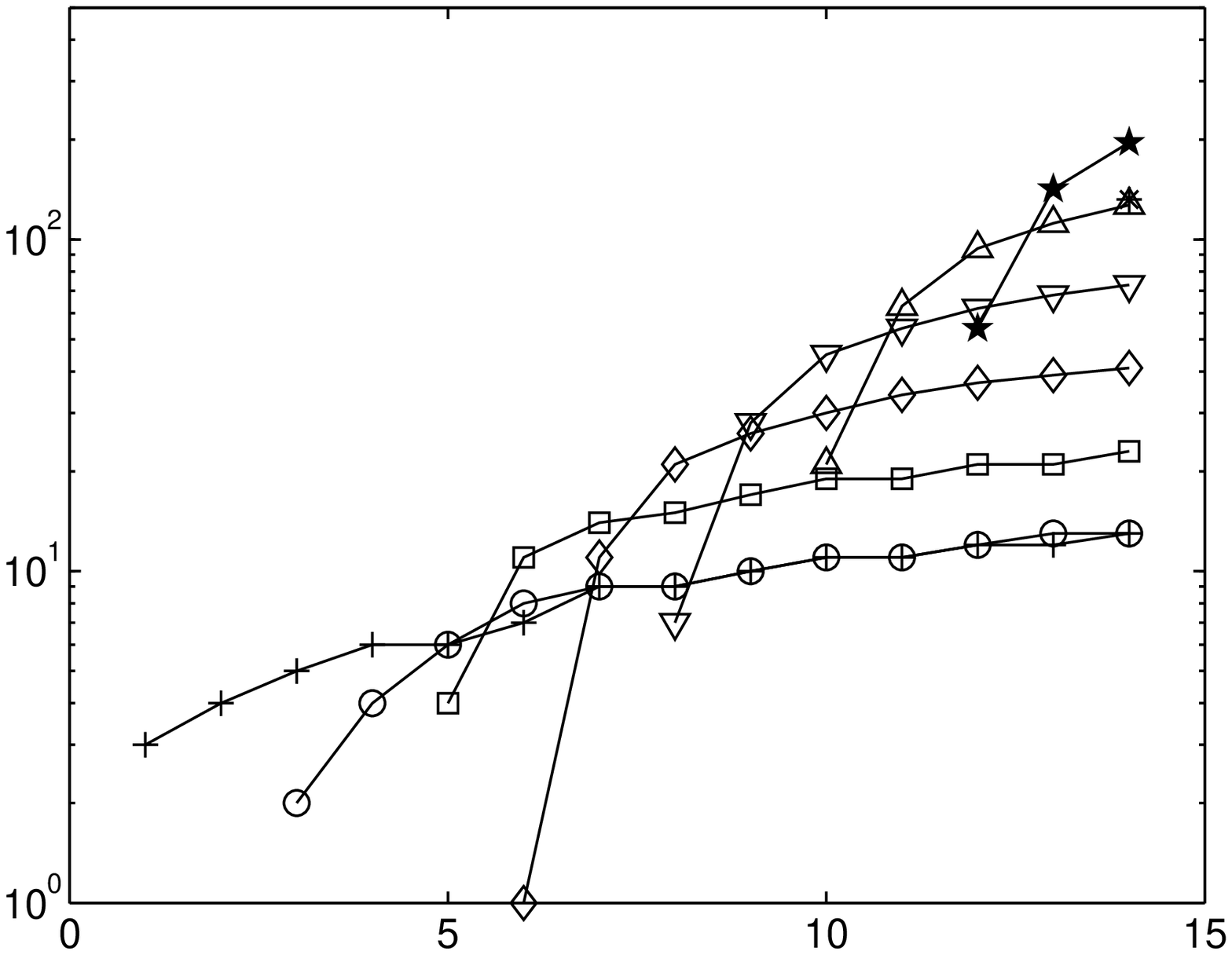}}%
    \put(0.539,0.02){\includegraphics[width=0.45\textwidth,height=0.35\textwidth]{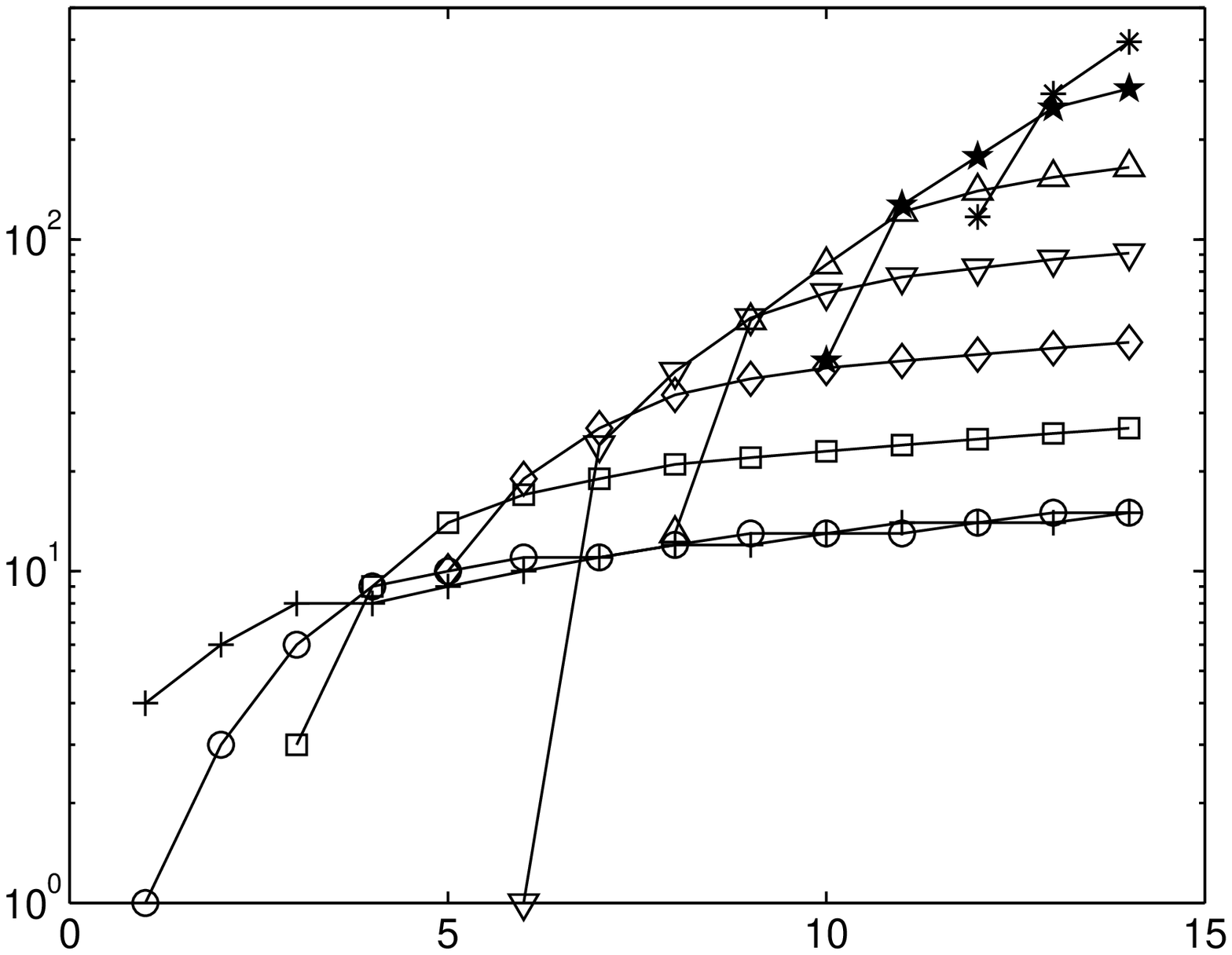}}%
    \put(0.039,0.42){\includegraphics[width=0.45\textwidth,height=0.35\textwidth]{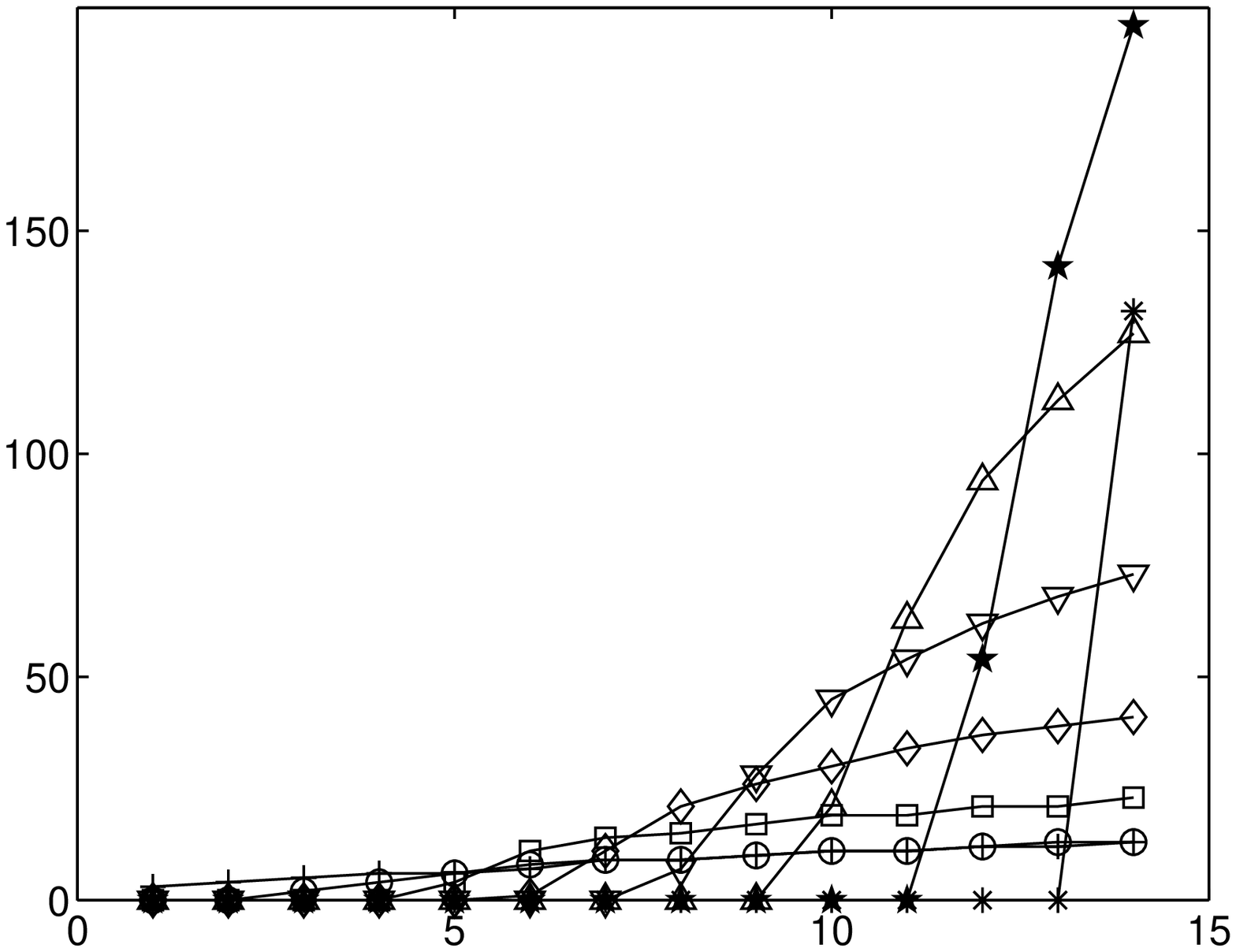}}%
    \put(0.539,0.42){\includegraphics[width=0.45\textwidth,height=0.35\textwidth]{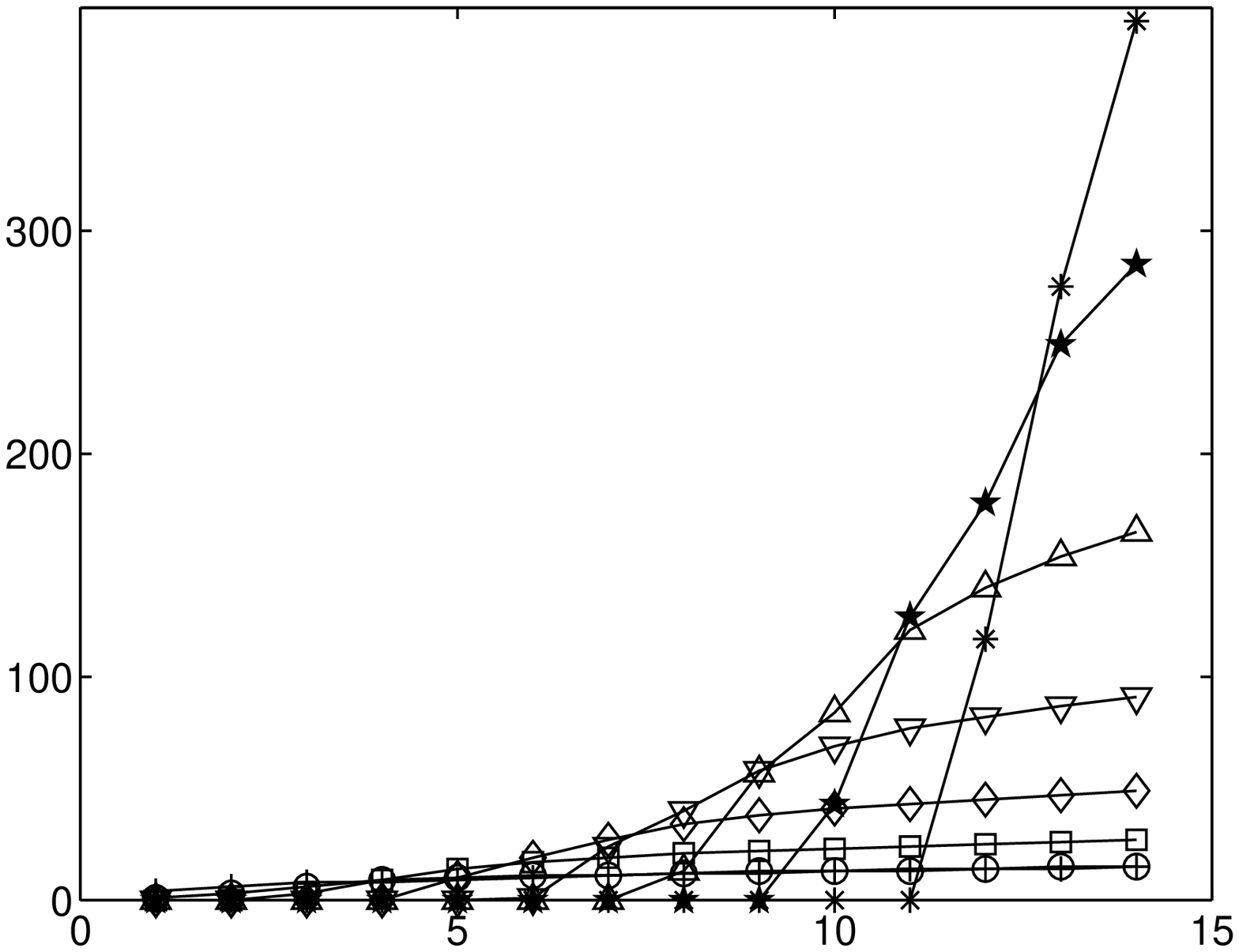}}%
    \put(-0.004,0.34){\small $\#\widetilde{W}_m$}
    \put(0.504,0.34){\small $\#\widetilde{W}_m$}
    \put(-0.004,0.74){\small $\#\widetilde{W}_m$}
    \put(0.504,0.74){\small $\#\widetilde{W}_m$}
    \put(0.109,0.70){\small $i=0$}
    \put(0.609,0.70){\small $i=3$}
  \end{picture}
\caption{\label{fig:x} The number of the significant basis functions in various subspaces
required to reproduce the $\|P_0\Psi_i\|^2$ and $\|Q_m\Psi_i\|^2$ to a given precision. The
horizontal axis shows the number of digit 9 in the threshold value $\eta$. The sign $+$ stands
for the restricted scaling function subspace $\widetilde{V}_0$, whereas the signs $\circ$,
\drawbox{1ex}{1ex}$\,$, $\lozenge$, $\triangledown$, $\vartriangle$, $\bigstar$ and $\ast$
denote the detail spaces $\widetilde{W}_m$, with $m=0,1,2,3,4,5,6$, respectively. The results
for the ground state ($i=0$) are plotted in the left column both in linear and log scale,
while the right column corresponds to the third excited state.}
\end{figure*}

After numerically calculating the scalar products
(\ref{expcoeff}), a threshold value $\eta$, close to 1 was chosen.
We counted the number $\#\widetilde{V}_0$ and $\#\widetilde{W}_m$
of most significant coefficients for which the inequalities
\begin{eqnarray}\label{THResh}
  \|P_0\Psi_i\|^2-\|\widetilde{P}_0\Psi_i\|^2 &\le& 1-\eta \qquad\mbox{and} \nonumber\\
  \|Q_m\Psi_i\|^2-\|\widetilde{Q}_m\Psi_i\|^2 &\le& 1-\eta
\end{eqnarray}
hold. This criterion allows each subspace $W_m$ to introduce an
error $1-\eta$ uniformly, and for those $m$ where
$\|Q_m\Psi_i\|^2\leq 1-\eta$ the complete subspace can be omitted,
thus $\widetilde{W}_m=\emptyset$. Fig.~\ref{fig:x} summarizes the
results for the ground and a selected excited state as a function
of $\eta$ for various resolution levels. On the horizontal axes
the number of digit 9 in $\eta$ is shown, i.e., the value 1
corresponds to $\eta=0.9$, whereas 5 corresponds to
$\eta=0.99999$, etc.
It is clearly seen, that ground and excited states behave in a very similar manner. As the
logarithmic plots show, in the $\eta\to1$ limit the number of basis functions in the subspaces
$W_m$ increases as $\sim 2^m$. On the other hand, if a finite, but even relatively high
precision, like $\eta=0.9999999999$ is demanded, many of the high resolution detail spaces
drop out, and even at lower resolution $m$, only an acceptable amount of basis functions
contribute to the truncated subspaces.

\section{Trajectory and limiting behavior of the eigenfunctions
in coarse-grained Hilbert spaces}

In applied calculations the granularity $M$ of the Hilbert space
(i.e., the maximum resolution $m$ in the summations of
(\ref{psidecomp}) and (\ref{Parseval})) can not be arbitrary
large. Instead, for a finite granularity $M$ the eigenvalue
problem is solved in the restricted Hilbert space
\begin{equation}\label{VWdecompM}
  \mathcal{H}^{[M]}=V_0\oplus W_0\oplus W_1\oplus\cdots\oplus W_{M-1},
\end{equation}
which assumption seems to be warranted in the light of the above
results. For $M=0$, we define $\mathcal{H}^{[0]}=V_0$. The
solution of the eigenvalue problem of $H$ in the coarse-grained
Hilbert space $\mathcal{H}^{[M]}$ leads to the ground and excited
states $\Psi^{[M]}_i$ ($i=0,1,2,\ldots$). In this case we lose
fine details, thus the natural question arises, how the trajectory
$\Psi^{[0]}_i\to \Psi^{[1]}_i\to\Psi^{[2]}_i\to \cdots$ approaches
the limit $\Psi_i$.

We have determined the eigenfunctions of the Hamiltonian (\ref{HO_Hamiltonian}) with
$\omega=1$ for $M=0,\ldots,5$. In order to keep the number of basis functions finite, the
spatial extension of the system has to be limited. We have chosen $|x|\leq
x_{\mathrm{max}}=10$ in (\ref{HO_Hamiltonian}). The error made by this assumption can be
estimated by the omitted part of the norm square of the ground state
\[
1-\int_{-x_\mathrm{max}}^{x_\mathrm{max}}|\Psi_0|^2\,
dx=1-\mathrm{erf}(\sqrt{\omega}\, x_\mathrm{max})
\]
which is less than the accuracy of the double precision numerical representation.

The four-index matrix elements of $H$ can be reduced to two-index quantities by a simple
variable transformation. The kinetic energy matrix elements are calculated using the fact
that the momentum operator is self-adjoint, thus for $m_1\leq m_2$
\begin{eqnarray}\label{KINmx}
  \langle s_{0\ell_1}|-\nabla^2/2\,|s_{0\ell_2}\rangle&=&
    2^{-1}\langle s'_{00}|s'_{0\,\ell_2-\ell_1}\rangle, \nonumber \\
  \langle s_{0\ell_1}|-\nabla^2/2\,|w_{m_2\ell_2}\rangle&=&
    2^{-1}\langle s'_{00}|w'_{m_2\,L}\rangle,                       \\
  \langle w_{m_1\ell_1}|-\nabla^2/2\,|w_{m_2\ell_2}\rangle&=&
    2^{2m_1-1}\langle w'_{00}|w'_{m_2-m_1\,L}\rangle,        \nonumber
\end{eqnarray}
with $L=\ell_2-2^{m_2-m_1}\ell_1$. The potential energy matrix
elements are
\begin{eqnarray}\label{POTmx}
  \langle s_{0\ell_1}|\omega^2x^2/2\,|s_{0\ell_2}\rangle&=&
        \omega^2 2^{-1}\bigl(\langle s_{00}|x^2|s_{0\,\ell_2-\ell_1}\rangle \nonumber \\
        &&\qquad+2\ell_1\langle s_{00}|x|s_{0\,\ell_2-\ell_1}\rangle        \nonumber \\
        &&\qquad+\ell_1^2\langle s_{00}|s_{0\,\ell_2-\ell_1}\rangle\bigr),  \nonumber \\
  \langle s_{0\ell_1}|\omega^2x^2/2\,|w_{m_2\ell_2}\rangle&=&
        \omega^2 2^{-1}\bigl(\langle s_{00}|x^2|w_{m_2\,L}\rangle           \nonumber \\
        &&\qquad+2\ell_1\langle s_{00}|x|w_{m_2\,L}\rangle                  \nonumber \\
        &&\qquad+\ell_1^2 \langle s_{00}|w_{m_2\,L}\rangle\bigr),                     \\
  \langle w_{m_1\ell_1}|\omega^2x^2/2\,|w_{m_2\ell_2}\rangle&=&
        \omega^2 2^{-2m_1-1}\cdot                                           \nonumber \\
        &&\qquad\bigl(\langle w_{00}|x^2|w_{m_2-m_1\,L}\rangle              \nonumber \\
        &&\qquad+2\ell_1\langle w_{00}|x|w_{m_2-m_1\,L}\rangle              \nonumber \\
        &&\qquad+\ell_1^2 \langle w_{00}|w_{m_2-m_1\,L}\rangle\bigr).       \nonumber
\end{eqnarray}
At any resolution level, the $x^2$ and $x$ operators can be exactly expanded as linear
combinations of scaling functions \cite{Daub}. Wavelets are also finite linear combinations of
scaling functions of the succeeding resolution level. Consequently, the remaining part of the
calculation contains the evaluation of integrals of type $\int s_{\mu_1\lambda_1}(x)
s_{\mu_2\lambda_2}(x) s_{\mu_3\lambda_3}(x)\,dx$ with $s_{\mu\lambda}(x)=2^{\mu/2}s_0(2^\mu
x-\lambda)$. There exist special algorithms for this task, and for exactly determining the
integrals of products of scaling function derivatives \cite{DahMicc}. Finally, the Hamiltonian
matrix was diagonalized by standard subroutines.

At each refinement level (granularity) $M$ the diagonalization of
the Hamiltonian matrix $H\big{|}\mbox{}_{\mathcal{H}^{[M]}}$
restricted to the appropriate subspace $\mathcal{H}^{[M]}$ leads
to the eigenvector components $c_{0\ell}^{[M]}$ and
$d_{m\ell}^{[M]}$, which are the expansion coefficients of
\begin{equation}\label{psiMdecomp}
  \Psi^{[M]}(x)=\sum_{\ell\in\mathbb{Z}} c_\ell^{[M]}\; s_{0 \ell}(x)+
                \sum_{\rule{0pt}{1.11ex}m=0}^{M-1}\sum_{\ell\in\Z}
                d_{m\ell}^{[M]}\;w_{m\ell}(x)
\end{equation}
for the approximate eigenstates. As the basis set of
$\mathcal{H}^{[M]}=\mathcal{H}^{[M-1]}\oplus W_{M-1}$ is an
extension to that of $\mathcal{H}^{[M-1]}$, the expansion
coefficients resulted from independent diagonalizations of
$H\big{|}\mbox{}_{\mathcal{H}^{[M-1]}}$ and
$H\big{|}\mbox{}_{\mathcal{H}^{[M]}}$ can change in the refinement
step $M-1\to M$. Describing the trajectory of $\Psi^{[M]}$ in
$\mathcal{H}$ in terms of the granularity, would need to trace all
the coefficients $c_{0\ell}^{[M]}$ and $d_{m\ell}^{[M]}$, this is
however, hard to demonstrate. Instead, we have collected the
ground state projections to a given detail space, and presented
them in Tab.~\ref{tab:egy}.
\begin{turnpage}
\begin{table*}
\caption{\label{tab:egy} The trajectory of the ground state wave function in the detail spaces
$V_0$, $W_0,\ldots,W_4$: the contribution of the detail spaces to the $M$th coarse-grained
approximations of the ground state eigenfunction. The last column shows the sum of the first
five terms in Parseval's equation (\ref{Parseval}). In finite subspaces $\mathcal{H}^{[0]},
\ldots, \mathcal{H}^{[5]}$ the relation fulfills exactly, whereas in the infinitely fine
Hilbert space the error is of order $10^{-12}$.}
\bigskip
\begin{tabular}{c@{$\quad$}c@{$\quad$}c@{$\quad$}c@{$\quad$}c@{$\quad$}c@{$\quad$}c@{$\quad$}c@{$\quad$}c}
     \hline\hline
     $M$ &$\bigl\|P_0\Psi^{[M]}_0\bigr\|^2$
         &$\bigl\|Q_0\Psi^{[M]}_0\bigr\|^2$
         &$\bigl\|Q_1\Psi^{[M]}_0\bigr\|^2$
         &$\bigl\|Q_2\Psi^{[M]}_0\bigr\|^2$
         &$\bigl\|Q_3\Psi^{[M]}_0\bigr\|^2$
         &$\bigl\|Q_4\Psi^{[M]}_0\bigr\|^2$
         &
         & Sum \\
     \hline
       0 & 1          &                         &                        &                         &                        &                       &      &  1 \\
       1 & 0.997552283& $2.4477165\times10^{-3}$&                        &                         &                        &                       &      &  1 \\
       2 & 0.997257786& $2.6805076\times10^{-3}$& $6.170620\times10^{-5}$&                         &                        &                       &      &  1 \\
       3 & 0.997234805& $2.7016396\times10^{-3}$& $6.247610\times10^{-5}$& $1.0791337\times10^{-6}$&                        &                       &      &  1 \\
       4 & 0.997233338& $2.7030347\times10^{-3}$& $6.252997\times10^{-5}$& $1.0801295\times10^{-6}$& $1.731278\times10^{-8}$&                       &      &  1 \\
       5 & 0.997233246& $2.7031230\times10^{-3}$& $6.253341\times10^{-5}$& $1.0801932\times10^{-6}$& $1.731382\times10^{-8}$& $2.7227\times10^{-10}$&      &  1 \\
     \hline
$\infty$ & 0.997233239& $2.7031289\times10^{-3}$& $6.253364\times10^{-5}$& $1.0801974\times10^{-6}$& $1.731389\times10^{-8}$& $2.7227\times10^{-10}$&\ldots& $1-4.34\times10^{-12}$\\
     \hline\hline
\end{tabular}
\end{table*}
\end{turnpage}
The excellent convergence can be easily realized. After few
refinement steps the coefficients stabilize.

For higher excitations the overall trend is similar to that
illustrated in Tab.~\ref{tab:egy} for the ground state, with the
remark, that the contributions of finer detail spaces are more and
more emphasized for increasing excitation levels. As an
illustration, we recall the projections of the $i=3$ excited state
to the wavelet subspaces at resolution level $M=5$:
$\|P_0\Psi^{[5]}_3\|^2=0.86367$, $\|Q_0\Psi^{[5]}_3\|^2=0.13284$,
$\|Q_1\Psi^{[5]}_3\|^2=3.4188\times10^{-3}$,
$\|Q_2\Psi^{[5]}_3\|^2=6.5664\times10^{-5}$,
$\|Q_3\Psi^{[5]}_3\|^2=1.0810\times10^{-6}$ and
$\|Q_4\Psi^{[5]}_3\|^2=1.7114\times10^{-8}$.

\begin{figure}[h!]
  \setlength{\unitlength}{\textwidth}
  \begin{picture}(0.47,0.39)
    \put(0.009,0.02){\includegraphics[width=0.45\textwidth,height=0.35\textwidth]{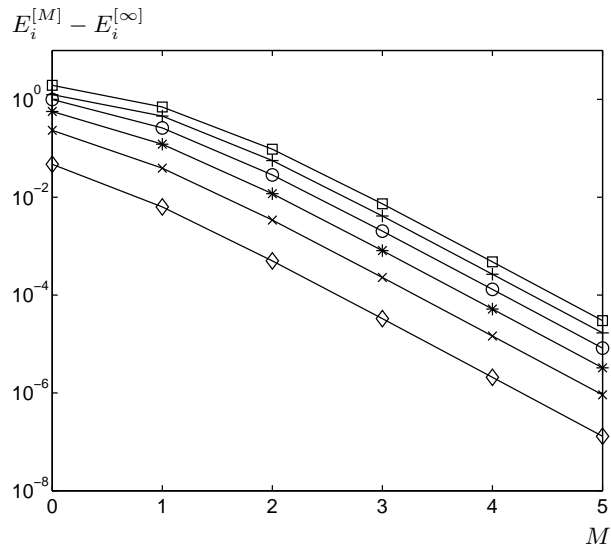}}%
    \put(0.009,0.38){\small $E_i^{[M]}-E_i^{[\infty]}$}
    \put(0.434,0.00){\small $M$}
  \end{picture}
\caption{\label{fig:y} The errors of the approximate energies $E_i^{[M]}$ determined in the
Hilbert space $\mathcal{H}^{[M]}$. The sign $\lozenge$ stands for the ground state $i=0$,
whereas the signs $\times$, $\ast$, $\circ$, $+$ and \drawbox{1ex}{1ex} denote the excitations
$i=1,2,3,4,5$, respectively. Atomic units are used.}
\end{figure}

For completeness, we have also given the errors of the ground and excited state energy
approximations in the Hilbert spaces $\mathcal{H}^{[M]}$ in Fig.~\ref{fig:y}. Exponential
convergence in terms of the granularity level $M$ can be clearly identified.

\section{An adaptive method for determining fine structure localization regions}

In realistic cases, of course, exact solutions of the eigenvalue
problem are not known. We can suppose, however, that quantum
mechanical systems  behave similarly to the simple exact case
studied above. We expect, that high resolution wavelets can
completely be omitted, and even at lower resolutions only a
fractional part of them contribute essentially to the expansion.
In the following considerations we will apply a consistent
notation for the two distinct truncation schemes. Upper index
$[M]$ denotes the ``vertical'' cutoff of the Hilbert space above
the granularity $M$, whereas tilde is applied to indicate the
``horizontal'' truncation in a given detail space $W_m$.

Our goal is to develop a method for determining the significant
coefficients in (\ref{psidecomp}), without calculating all the
negligible ones. This requires predicting the set of important
fine level coefficients for $M+1$, supposing that the approximate
expansion
\begin{equation}\label{Psiapproxm}
\widetilde{\Psi}^{[M]}(x)=\sum_{\ell\in \widetilde{V}_0}
\tilde{c}^{[M]}_\ell\; s_{0
\ell}(x)+\sum_{\rule{0pt}{1.58ex}m=0}^{M-1}\sum_{\ell\in\widetilde{W}_m}\tilde{d}^{[M]}_{m
\ell}
  \;w_{m\ell}(x).
\end{equation}
of the wave function is already known. Here, we have used the
loose notation $\ell\in\widetilde{W}_m$, indicating the fact, that
the summation index $\ell$ is constrained to the wavelets
$w_{m\ell}$ which belong to the restricted subspace
$\widetilde{W}_m$. In the first summation $\widetilde{V}_0$
denotes the restricted subspace of those scaling functions which
are localized in the finite domain of the space, which is occupied
by the system.

The iterative extension
$\widetilde{\Psi}^{[M]}\to\widetilde{\Psi}^{[M+1]}$ can be
separated to two independent problems. The first step is
determining the significant part $\widetilde{W}_M$ of the $M$th
level detail space, knowing the approximation
$\widetilde{\Psi}^{[M]}$. In the other step, the numerical
determination of the expansion coefficients $\tilde{d}_{M\ell}$ in
$\widetilde{W}_M$ should be carried out. In this work, we will
concentrate on the solution of the first problem. In order to
distinguish the errors emerging from different sources, the second
part of the calculation is carried out by an exact diagonalization
of the Hamiltonian $\widetilde{H}^{[M+1]}$ restricted to the
subspace
$\widetilde{\mathcal{H}}^{[M+1]}=\widetilde{V}_0\oplus\widetilde{W}_0\oplus\cdots\oplus\widetilde{W}_M$.
In optimized numerical works sophisticated methods for solving the
eigenproblem of sparse matrices can be applied.

As $\widetilde{\Psi}^{[M]}$ is the eigenfunction of the restricted
Hamiltonian
$\widetilde{H}^{[M]}=H\big|\mbox{}_{\widetilde{\mathcal{H}}^{[M]}}$,
it satisfies the equation
\begin{equation}\label{restr-eigeq}
 (\widetilde{H}^{[M]}-\widetilde{E}^{[M]})\widetilde{\Psi}^{[M]}= 0.
\end{equation}
This condition does not hold, if either $\widetilde{H}^{[M]}$ or
$\widetilde{E}^{[M]}$ are replaced by their exact counterparts,
and the quality of the $M$th resolution level approximate solution
of the eigenvalue problem can be characterized by the error
function
\begin{equation}\label{errorfunc}
 (H-\widetilde{E}^{[M]})\widetilde{\Psi}^{[M]}\neq 0.
\end{equation}
Based on the methods developed in \cite{Ladanyi,Dahmen}, we
suggest the following criterion for selecting the important basis
functions of $W_M$. The error function (\ref{errorfunc}) will be
measured in the extended Hilbert space $\mathcal{H}^{[M+1]}$
(instead of the full Hilbert space $\mathcal{H}$). Considering,
that according to (\ref{restr-eigeq}) the expansion coefficients
of the error function $\langle w_{m\ell}|
(H-\widetilde{E}^{[M]})|\widetilde{\Psi}^{[M]}\rangle=0$ if
$w_{m\ell}\in \widetilde{\mathcal{H}}^{[M]}$, the magnitude of the
deviation from the exact eigenvalue problem can be characterized
by the magnitude of the expansion coefficients in the detail space
$W_M$
\begin{equation}\label{errorabs}
r_{M\ell}=\bigl|\langle
w_{M\ell}|H-\widetilde{E}^{[M]}|\widetilde{\Psi}^{[M]}\rangle\bigr|^2.
\end{equation}
The wavelet $w_{M\ell}$ is considered to be ``important'' if
$r_{M\ell}$ is larger then a given threshold. This test, however,
can not be carried out in a reasonable time for all the basis
functions in $W_M$, as this would require an exponential amount of
work. Here, we apply an \emph{adaptive} approach. We test the
wavelets $w_{M\ell}$ only for those values of $\ell$, where the
support of $w_{M\ell}$ overlaps with the previously found
``important region'' of $W_{M-1}$.

\noindent To be more specific, we apply the following procedure.
\begin{enumerate}
 \item We select a preliminary set $\widehat{W}_M$ of wavelets $w_{M\ell}$ for
 which the condition
 \[
  \mathrm{support} \bigl(w_{M\ell}\bigr)\cap\;\mathrm{support}
  \bigl(\widetilde{W}_{M-1}\bigr)\neq\emptyset
 \]
 holds.
 \item For the selected wavelets the values of (\ref{errorabs})
 are calculated and the wavelets are sorted according to descending
 order of $r_{M\ell}$.
 \item The first most important wavelets constituting the restricted subspace $\widetilde{W}_M$ are kept until the
 condition
 \begin{equation}\label{criterion}
  \sum_{\ell\in{\widehat{W}_M}}r_{M\ell}-
  \sum_{\ell\in{\widetilde{W}_M}}r_{M\ell}
  \leq 1- \eta,
 \end{equation}
similar to (\ref{THResh}) is fulfilled. Depending on the required
precision $\eta$ the set $\widetilde{W}_M$ is usually much smaller
then the candidate set $\widehat{W}_M$.
\end{enumerate}
The adaptive fine structure localization method reduces the size
of the detail spaces considerably, for example, in the case of
nine-9 precision ($\eta=0.999999999$) and the ground state wave
function, $\dim W_6=646$, whereas $\dim\widetilde{W}_6=296$.

The results obtained for the wave function with various precision
requirements $\eta$ are summarized in Tables \ref{tab:eta99} and
\ref{tab:eta999999}. The wave function $\widetilde{\Psi}^{[M]}$
calculated in the ``horizontally'' truncated Hilbert space is
expected to significantly deviate from the eigenstate $\Psi^{[M]}$
received without truncation if the required precision is low. On
the other hand, in the $\eta\to1$ limit $\widetilde{\Psi}^{[M]}$
approximates $\Psi^{[M]}$ very well.

%\begin{turnpage}
\begin{table*}[t!]
\caption{\label{tab:eta99} The deviation of the trajectory of $\widetilde{\Psi}^{[M]}_0$ from
that of $\Psi^{[M]}_0$ due to fine structure localization with $\eta=0.99$ in the detail
spaces $V_0$, $W_0,\ldots,W_4$.}
\bigskip
\begin{tabular}{c@{$\quad$}c@{$\quad$}c@{$\quad$}c@{$\quad$}c@{$\quad$}c@{$\quad$}c}
     \hline\hline
     $M$ &$\Delta\bigl\|P_0\widetilde{\Psi}^{[M]}_0\bigr\|^2$
         &$\Delta\bigl\|Q_0\widetilde{\Psi}^{[M]}_0\bigr\|^2$
         &$\Delta\bigl\|Q_1\widetilde{\Psi}^{[M]}_0\bigr\|^2$
         &$\Delta\bigl\|Q_2\widetilde{\Psi}^{[M]}_0\bigr\|^2$
         &$\Delta\bigl\|Q_3\widetilde{\Psi}^{[M]}_0\bigr\|^2$
         &$\Delta\bigl\|Q_4\widetilde{\Psi}^{[M]}_0\bigr\|^2$
         \\
     \hline
       1 & $0.41770\times10^{-5}$& $0.41770\times10^{-5}$&                          &                         &                        &                        \\
       2 & $3.22557\times10^{-5}$& $3.22325\times10^{-5}$& $-0.23237\times10^{-7}$&                         &                        &                        \\
       3 & $3.14686\times10^{-5}$& $3.17587\times10^{-5}$& $\ \ \;2.50837\times10^{-7}$& $3.92755\times10^{-8}$&                        &                        \\
       4 & $2.76469\times10^{-5}$& $2.82869\times10^{-5}$& $\ \ \;5.95967\times10^{-7}$& $4.31415\times10^{-8}$& $0.89745\times10^{-9}$&                        \\
       5 & $2.75634\times10^{-5}$& $2.81104\times10^{-5}$& $\ \ \;5.11324\times10^{-7}$& $3.44685\times10^{-8}$& $1.19115\times10^{-9}$& $2.901\times10^{-11}$  \\
     \hline\hline
\end{tabular}
\end{table*}
%\end{turnpage}

Tab.~\ref{tab:eta99} contains the deviations
$\Delta\|P_0\widetilde{\Psi}_0^{[M]}\|^2=\|P_0\widetilde{\Psi}_0^{[M]}\|^2-\|P_0\Psi_0^{[M]}\|^2$
and
$\Delta\|Q_m\widetilde{\Psi}_0^{[M]}\|^2=\|Q_m\widetilde{\Psi}_0^{[M]}\|^2-\|Q_m\Psi_0^{[M]}\|^2$
for a rough precision $\eta=0.99$. It can be seen, that the error
of the wave function saturates with increasing resolution level
$M$. The value of the error is significantly larger than the error
caused by the ``vertical'' truncation of the Hilbert space.

%\begin{turnpage}
\begin{table*}
\caption{\label{tab:eta999999} The deviation of the trajectory of $\widetilde{\Psi}^{[M]}_0$
from that of $\Psi^{[M]}_0$ due to fine structure localization with $\eta=0.999999$ in the
detail spaces $V_0$, $W_0,\ldots,W_4$.}
\bigskip
\begin{tabular}{c@{$\quad$}c@{$\quad$}c@{$\quad$}c@{$\quad$}c@{$\quad$}c@{$\quad$}c}
     \hline\hline
     $M$ &$\Delta\bigl\|P_0\widetilde{\Psi}^{[M]}_0\bigr\|^2$
         &$\Delta\bigl\|Q_0\widetilde{\Psi}^{[M]}_0\bigr\|^2$
         &$\Delta\bigl\|Q_1\widetilde{\Psi}^{[M]}_0\bigr\|^2$
         &$\Delta\bigl\|Q_2\widetilde{\Psi}^{[M]}_0\bigr\|^2$
         &$\Delta\bigl\|Q_3\widetilde{\Psi}^{[M]}_0\bigr\|^2$
         &$\Delta\bigl\|Q_4\widetilde{\Psi}^{[M]}_0\bigr\|^2$
         \\
     \hline
       1 & $     -0.00209\times10^{-9}$& $\ \ \;0.00209\times10^{-9}$&                             &                      &                          &             \\
       2 & $\ \ \;0.21649\times10^{-9}$& $     -0.21870\times10^{-9}$& $\ \ \;0.0220\times10^{-10}$&                      &                          &             \\
       3 & $     -1.02284\times10^{-9}$& $\ \ \;1.04319\times10^{-9}$& $     -0.1999\times10^{-10}$& $-0.036\times10^{-11}$&                          &             \\
       4 & $\ \ \;4.15413\times10^{-9}$& $     -4.42360\times10^{-9}$& $\ \ \;2.9803\times10^{-10}$& $-2.845\times10^{-11}$& $     -1.0\times10^{-13}$&             \\
       5 & $\ \ \;3.26705\times10^{-9}$& $     -3.50721\times10^{-9}$& $\ \ \;2.6070\times10^{-10}$& $-2.072\times10^{-11}$& $\ \ \;1.8\times10^{-13}$& $-10^{-14}$  \\
     \hline\hline
\end{tabular}
\end{table*}
%\end{turnpage}

Tab.~\ref{tab:eta999999} shows similar data for a moderate
precision requirement $\eta=0.999999$. In this case the wave
function $\widetilde{\Psi}^{[M]}$, obtained using the adaptive
fine structure localization method, gives an excellent
approximation to $\Psi^{[M]}$. The error introduced by the
``horizontal'' truncation is less, than that of the ``vertical''
one, up to the resolution level $M=5$.

In case of excited states, the eigenstates of the eigenvalue
problem in granular Hilbert spaces with a given granularity $M$
are less accurate than the ground state. The deviations due to
``horizontal'' truncation of the detail spaces, however, only
slightly exceed the ground state values. Consequently, for a given
$M$, the precision $\eta$ which was appropriate for ground state
calculations will certainly be applicable in excited state
calculations, as well.

\begin{figure*}[t!]
  \setlength{\unitlength}{\textwidth}
  \begin{picture}(1,0.45)
    \put(0.015,0.02){\includegraphics[width=0.45\textwidth,height=0.35\textwidth]{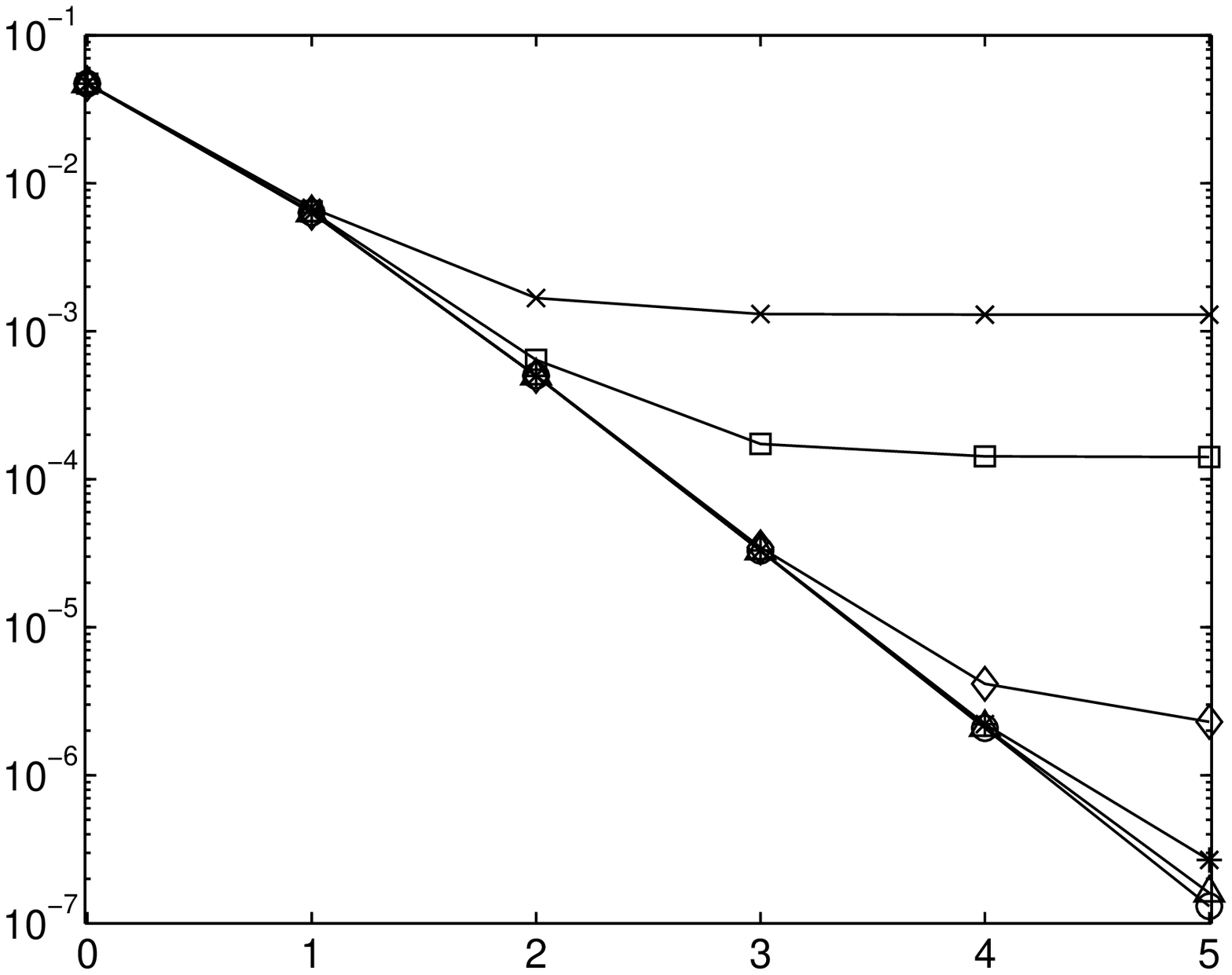}}%
    \put(0.515,0.02){\includegraphics[width=0.45\textwidth,height=0.35\textwidth]{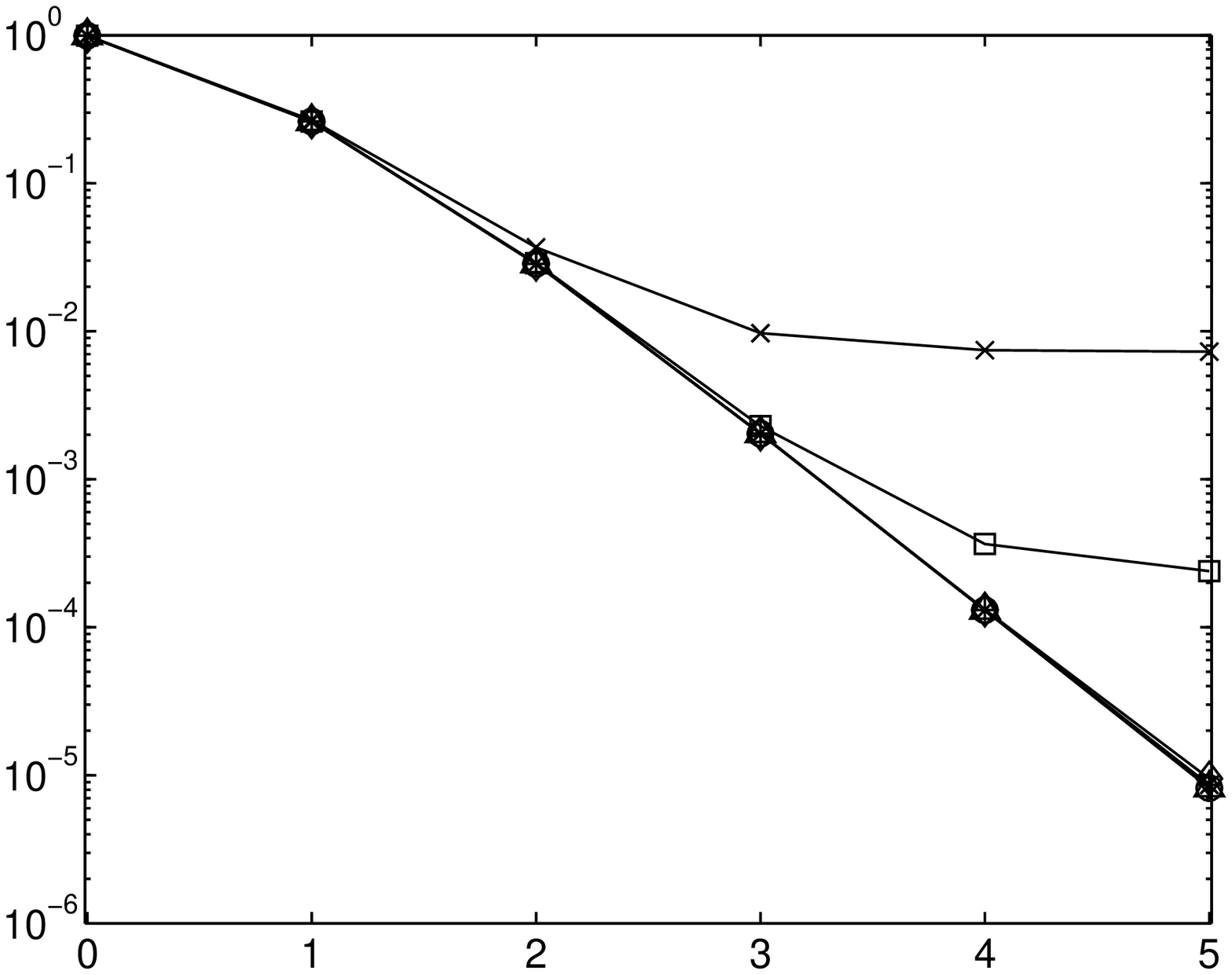}}%
    \put(0.015,0.38){\small $\widetilde{E}_i^{[M]}-E_i^{[\infty]}$}
    \put(0.515,0.38){\small $\widetilde{E}_i^{[M]}-E_i^{[\infty]}$}
    \put(0.115,0.15){\small $i=0$}
    \put(0.615,0.15){\small $i=3$}
    \put(0.440,0.00){\small $M$}
    \put(0.940,0.00){\small $M$}
  \end{picture}
\caption{\label{fig:Energ}The errors of the approximate energies $\widetilde{E}_i^{[M]}$
determined in the Hilbert space $\widetilde{\mathcal{H}}^{[M]}$ for the ground state $i=0$ and
for an excited state $i=3$. The signs $\times$, \drawbox{1ex}{1ex}$\,$, $\lozenge$, $\ast$,
$\triangle$ and $\circ$ stand for the threshold values $\eta=0.9,\ 0.99,\ 0.999,\ 0.9999,\
0.99999$ and 0.999999, respectively. Atomic units are used.}
\end{figure*}

The errors of the approximate energies calculated for the ground and excited state $i=3$
are plotted in Fig.~\ref{fig:Energ} for various precision requirements $\eta$. For lower
precision the magnitude of the error saturates at various resolution levels $M$, depending
on the value of $\eta$. This fact, together with the observation concerning the error of
the wave function, emphasizes the importance of choosing matching values for the precision
requirement $\eta$ and the granularity level $M$.

\section{Conclusions}

We have shown, that the fine structure of the Hilbert space
vanishes exponentially in realistic wave functions. Realizing this
fact, one can obtain excellent approximations of the
eigenfunctions by omitting the fine resolution detail spaces and
solving the eigenvalue problem in granular Hilbert spaces. We have
traced the trajectories of the approximate wave functions of an
exactly solvable model system, and concluded, that they approach
the exact wave function exponentially fast. A similar statement is
valid for the approximate energy values.

A further possibility for reducing the computational complexity of the calculations, that
higher resolution wavelets are included in the basis set only in those spatial regions
where the fine structure of the wave function requires it. For our model system we found,
that the number of significant basis functions is considerably less than the dimension of
the detail subspaces included in full calculations. The \emph{a priori} selection of the
significant basis functions needs a predictive algorithm.

Based on the above concept, we have developed an adaptive method for selecting the essential
basis functions using the fine structure localization technique. Calculations in truncated
Hilbert spaces, restricted in such manner lead to sufficiently precise wave functions and
eigenenergies, even in the case of moderately strict basis function selection criterion.

%\clearpage
\section*{ACKNOWLEDGMENTS}

This work was supported by the Orsz\'agos Tudom\'anyos Kutat\'asi
Alap (OTKA), Grant Nos. T046868 and NDF45172. %(T042981 ?????).
Fruitful discussions with Prof.\ S.~Dahlke and his research group are gratefully
acknowledged.

\end{document}